\documentclass[prl,twocolumn,superscriptaddress,showpacs]{revtex4-1}

\usepackage{amsmath}
\usepackage{chemarr}
\usepackage{graphicx}
\usepackage{url}
\usepackage{subfigure}

\begin{document}
%%%%%%%%%%%%%%%%%%%%%   The First Page    %%%%%%%%%%%%%%%%%%%%%%%%%%%%%

  \title{Potential and Flux Decomposition for Dynamical Systems and Non-Equilibrium Thermodynamics: Curvature, Gauge Field and Generalized Fluctuation-Dissipation Theorem}
  \date{\today}
  \author{Haidong Feng}
  \affiliation{Department of Chemistry, Physics, and Applied Mathematics, State University of New York at Stony Brook}
  \author{Jin Wang} \email[Corresponding author:]{jin.wang.1@stonybrook.edu}
  %\affiliation{Department of Chemistry, State University of New York at Stony Brook}
  %\affiliation{Department of Chemistry, State University of New York at Stony Brook}
  %\affiliation{}{1}
  \affiliation{State Key Laboratory of Electroanalytical Chemistry, Changchun Institute of Applied Chemistry, Chinese Academy of Sciences}

\begin{abstract}

The driving force of the dynamical system can be decomposed into the
gradient of a potential landscape and curl flux (current). The
fluctuation-dissipation theorem (FDT) is often applied to near
equilibrium systems with detailed balance. The response due to a
small perturbation can be expressed by a spontaneous fluctuation.
For non-equilibrium systems, we derived a generalized FDT that the
response function is composed of two parts: (1) a spontaneous
correlation representing the relaxation which is present in the near
equilibrium systems with detailed balance; (2) a correlation related
to the persistence of the curl flux in steady state, which is also
in part linked to a internal curvature of a gauge field. The
generalized FDT is also related to the fluctuation theorem. In the
equal time limit, the generalized FDT naturally leads to
non-equilibrium thermodynamics where the entropy production rate can
be decomposed into spontaneous relaxation driven by gradient force
and house keeping contribution driven by the non-zero flux that
sustains the non-equilibrium environment and breaks the detailed
balance.
\end{abstract}

\maketitle

The global stability is essential in understanding the dynamical
non-equilibrium systems. The driving force of the dynamical system
often is not integrable and can not be written in terms of the
gradient of a potential. The driving force however can be decomposed
into the gradient of a potential and a curl flux (current)
\cite{Wang}. The potential is related to the steady state
probability and the gradient force gives the normal dynamics
analogous to equilibrium system, while the curl flux force
is directly linked to the non-equilibrium contribution from detailed
balance breaking. For non-equilibrium dynamics, the dual
description with both potential and flux is necessary.

In addition, the fluctuation-dissipation theorem (FDT) plays a
central role for systems in near equilibrium systems with detail
balance \cite{Kubo, Deker}. It links the fluctuations of the system
quantified by the correlation function with the response of the
system quantified by the response function. Many efforts have been
made to extend the FDT to non-equilibrium systems \cite{Cugliandolo,Hanggi,
Chetrite,Hatano, Verley, Baiesi, Zannetti, Vulpiani, Okabe,
Lu, Gawedzki, Bechinger, Maes}. It was found that the FDT  involves
the correlation function of a variable that is conjugate with
entropy \cite{Seifert}. Furthermore, by choosing proper
observables, the FDT for non-equilibrium systems can be uncovered
\cite{Prost}.
%But the physical mean of such observables is not so
%straight forward in non-equilibrium cases.
%But such observables are often not direct observables
%in the experiments as in the FDT of equilibrium cases.

In this letter, we found another way to generalize FDT for
non-equilibrium processes, specifically for direct observables such
as $x_i$, under Markov dynamics in continuous space described by
{\it Langevin dynamics} or {\it Fokker-Planck equations}.
Particularly, the response function can be split into two parts. One
is from the correlation of the observable itself representing the
spontaneous relaxations,
%which links with the gradient part of the force.
which also exists in systems with detailed balance. The other one
relates to the heat dissipation in the medium, representing the
detailed balance breaking contribution, which directly links to the
curl flux part of the force. On a closed loop, the medium heat
dissipation can be described by the internal curvature introduced by
the non-gradient force or curl flux part, which is analogous to {\it
Abelian Gauge Theory} \cite{Peskin}. On any particular path, the
medium heat dissipation is analogous to the {\it Wilson lines} of
{\it Abelian gauge theory} % in early literature of {\it QED}
\cite{Peskin}. In the equal time limit, the generalized FDT
naturally leads to non-equilibrium thermodynamics \cite{Ge,
Seifert1, Esposito}. In addition, this generalized FDT is also
related to {\it Fluctuation Theorem} \cite{Morriss, Searles, Evans,
Andrieux, Szabo, Jarzynski}.

Markov dynamics in continuous space can be characterized by {\it Langevin} equations:
\begin{eqnarray}\label{LE}
\dot{x}_i= F_i ({\bf x}) + B_{ij} ({\bf x}) \xi_j (t)
\end{eqnarray}
where $F_i ({\bf x}) $ is the driving force and $\xi_i (t)$ is the
 Gaussian distributed white noise:
$\langle  \xi_i (t)  \xi'_j (t') \rangle =   \delta (t-t')$.
Here the {\it Einstein notation} is used: when an index $i$ appears twice in a single term, it implies that we are summing over all of its possible values.
The probability obeys the {\it Fokker-Planck equation}:
\begin{eqnarray}\label{FPE}
\dot{P} ({\bf x},t) = \hat{L} ({\bf x}) P({\bf x},t)
\end{eqnarray}
with the operator $\hat{L} ({\bf x}) = \Big [
-\partial_i  F_i ({\bf x}) +
\partial_i \partial_j D_{ij} (\bf x) \Big ] $
and the diffusion coefficient $D_{ij} ({\bf x}) = \frac{1}{2} ({\bf
B B^T})_{ij} (\bf x)$. For convenience, we use $\partial_i \equiv \frac{\partial}{\partial x_i}$, $P ({\bf x}) \equiv
P({\bf x},t)$ to represent the time dependent probability
distribution and $P^{SS} ({\bf x})$ to indicate the time independent
steady state probability distribution.
%If the driving force is a gradient: $F_i ({\bf x}) = - \frac{\partial U({\bf x})}{\partial x_i}$,
The flux can be defined as:
%\begin{equation}
%-F_i ({\bf x}) P ({\bf x}) + \frac{\partial}{\partial x_j} D_{ij}
%({\bf x}) P ({\bf x}) = j_i ({\bf x})
%\end{equation}
%or equivalently
\begin{equation}\label{current1}
 - \tilde{F}_i ({\bf x}) P ({\bf x}) + D_{ij} ({\bf x}) \partial_j  P ({\bf x}) = j_i ({\bf x})
\end{equation}
where $\tilde{F}_i  = F_i - \partial_j D_{ij}$.
Then {\it Fokker-Planck equation} can be rewritten as $\frac{d P({\bf x},t)}{dt}= {\bf \partial \cdot j}$. The
system is considered to be in detailed balance if the steady state
flux:
\begin{equation}\label{distri_ndb}
- \tilde{F}_i ({\bf x}) P^{SS} ({\bf x}) + D_{ij} ({\bf x}) \partial_j P^{SS} ({\bf x}) = j^{SS}_i ({\bf x})
\end{equation}
is zero: ${\bf j}^{SS}=0$. For general non-equilibrium systems
without detailed balance: ${\bf j}^{SS} \neq 0$, the steady state
flux is a divergence free vector with ${\bf \partial \cdot
j}^{SS}=0$. The force term $\tilde{F}_j ({\bf x})$ can be decomposed
into two parts: a potential gradient term $-D_{ij} ({\bf x})
\frac{\partial}{\partial x_i} U ({\bf x})$ where $U({\bf x})=-\ln
P_{ss}({\bf x})$ and flux term $ - j^{SS}_j ({\bf x}) / P^{SS}
({\bf x}) \equiv - v_j ({\bf x})$, with a probabilistic velocity:
$v_i ({\bf x})$. Alternatively, the gradient of potential $-\ln
P^{SS}({\bf x})$ can also be decomposed into a force term and a curl
flux term :
\begin{equation}\label{current2}
 -\partial_i \ln [P^{SS}({\bf x})]
=D^{-1}_{ij} ({\bf x}) [- \tilde{F}_j ({\bf x})  -
v^{SS}_j ({\bf x})]
\end{equation}

%The correlation between two observables $\varOmega^1$ and
%$\varOmega^2$ can be defined as: $C_{{\varOmega}^1 {\varOmega}^2}
%(0, t) = \langle {\varOmega}^1 (0) {\varOmega}^2 (t) \rangle -
%\langle {\varOmega}^1 (0) \rangle \langle {\varOmega}^2 (t) \rangle$
%with $\langle {\varOmega}^1 (0) {\varOmega}^2 (t) \rangle =  P_{SS}
%({\bf x}^i)  \varOmega^1_{{\bf x}^i} \varOmega^2_{{\bf x}^j} P({\bf
%x}^i, 0| {\bf x}^j, t) $. $P({\bf x}^i, 0| {\bf x}^i, t)$ is the
%conditional transition probability from initial state ${\bf x}^i$ at
%time $t=0$ to final state ${\bf x}^j$ at time $t$.
%Upon a pulse-like perturbation $h(t)$, the response function (through observable ${\varOmega}$) of
%a system initially prepared in the steady state $P^0_{SS} (i)$
%is:
%$R^{{\varOmega}} (t_0,t) = \frac{\delta (\langle \Omega \rangle (t) - \langle \Omega \rangle^0)}{\delta h (t_0)} \Big \arrowvert_{h=0}$
%in which $\langle \Omega \rangle^0$ is the mean value of $\Omega$
%for the unperturbed steady state.

Using perturbation theories, FDT  for equilibrium systems with
detailed balance was investigated \cite{Deker}. Here we
will extend it to non-equilibrium systems.
% and find that the
%response function can be contributed by two part: one is the
%correlation function of the observables themselves representing
%the spontaneous relaxation and the other one relates to the steady
%state flux ${\bf j}$ defined in equ. (\ref{distri_ndb}).
Consider a linear perturbation on the force: $F_i ({\bf x})
\rightarrow F'_i ({\bf x}) = F_i ({\bf x}) + h (t) \delta F_i
({\bf x})$, we have
$\hat{L} \rightarrow \hat{L}' = \hat{L} -  h (t) \delta \hat{L}$,
%&=&  \hat{L} -  h (t) [ \delta F_i ({\bf x}) \frac{\partial}{\partial x_i} +
%\frac{\partial \delta  F_i ({\bf x})}{\partial x_i} ]
%\end{eqnarray}
%with
%\begin{equation}
%\delta \hat{L} = \delta F_i ({\bf x}) \frac{\partial}{\partial x_i} +
%\frac{\partial \delta  F_i ({\bf x})}{\partial x_i}
%\end{equation}
with $\delta \hat{L}= \delta F_i ({\bf x}) \partial_i +
 \partial_i \delta  F_i ({\bf x})$. The probability
evolves as %for operator $\Omega({\bf x})$
\begin{eqnarray}
& P({\bf x}, t) = exp \Big [ \int_{t'}^t dt ( \hat{L} -  h_i (t) \delta \hat{L} ) \Big] P ({\bf x}, t') \\
& \delta \langle \Omega(t) \rangle = \langle \Omega(t) \rangle - \langle \Omega \rangle = \int d {\bf x} \Omega({\bf x}) [P({\bf x}, t) - P^{SS}({\bf x})] \nonumber
\end{eqnarray}
Therefore, for $t \geq t'$, the response function reads as
\begin{eqnarray}
&& R^{\Omega}_i (t-t') = \delta \langle \Omega (t) \rangle / \delta h (t') \Big \arrowvert_{\delta {\bf F}=0} \nonumber \\
&=& \int d {\bf x} \Omega({\bf x}) e^{\hat{L} (t-t')} (- \delta \hat{L}) P^{SS}({\bf x})
\end{eqnarray}
%with the step function $\Theta (t)$.
Using the decomposition in equ. (\ref{current2}),
%By rewriting $j^{SS}_i = (j^{SS}_i/P^{SS}) P^{SS}$, for $t\geq t'$,
we have
\begin{eqnarray}\label{R}
&& R^{\Omega}_i (t-t') \\
&=&  \int d {\bf x} \Omega e^{\hat{L} (t-t')} \{\delta F_i [- \tilde{F}_k  - v^{SS}_k ] D^{-1}_{ik} -  \partial_i \delta  F_i \}
P^{SS} \nonumber \\
&= &- \langle \Omega(t) \partial_i \delta  F_i (t')\rangle -  \Big [ \langle  \Omega(t) \delta F_i (t') \tilde{F}_k (t') D^{-1}_{ik} (t') \rangle \nonumber \\
&& \quad \quad \quad \quad \quad \quad \quad \quad \quad + \langle \Omega(t) \delta F_i (t')  v^{SS}_k (t')  D^{-1}_{ik} (t') \rangle
\Big ] \nonumber
\end{eqnarray}
This is the general relation between response functions and correlation functions.
Here, the correlation between two observables $\varOmega^1$ and $\varOmega^2$ is
$C_{{\varOmega}^1 {\varOmega}^2} (t', t) = \langle {\varOmega}^1 (t') {\varOmega}^2 (t) \rangle
- \langle {\varOmega}^1 (t') \rangle \langle {\varOmega}^2 (t) \rangle$
with
$\langle {\varOmega}^1 (t') {\varOmega}^2 (t) \rangle =  P_{SS} ({\bf x}^i)  \varOmega^1_{{\bf x}^i} \varOmega^2_{{\bf x}^j}
P({\bf x}^i, t'| {\bf x}^j, t)$. $P({\bf x}^i, t'| {\bf x}^i, t)$ is the transition probability from
initial state ${\bf x}^i$ at time $t'$ to final state ${\bf x}^j$ at time $t$.
For the perturbation independent on ${\bf x}$: $\delta F_i =1$, we obtain  % and homogenous diffusion coefficient $D_{ij}= const$.
\begin{eqnarray}\label{R1}
&& R^{\Omega}_i (t-t') = - \langle \Omega(t) \partial_i \ln [P^{SS}({\bf x})] \rangle \\
& =& -\Big [ \langle  \Omega(t) \tilde{F}_k (t') D^{-1}_{ik} (t') \rangle
+ \langle \Omega(t)  v^{SS}_k (t') D^{-1}_{ik} (t') \rangle  \nonumber
\Big ]
\end{eqnarray}
which is a generalized FDT for non-equilibrium systems
\cite{Seifert1}. With the force decomposition in equ.
(\ref{current2}), the response of the system is composed of two
terms. The first term, just as equilibrium cases, is related to the
usual correlation of the variable with the driving force. This term
exists even for FDT of equilibrium systems obeying the detailed
balance (this is the case where the gradient of the logarithm of
probability is equal to the driving force). The second term however
is directly related to the non-zero flux which violates the detailed
balance and measures the degree of the non-equilibrium-ness (how far
away the system is from equilibrium).

FDT in equ. (\ref{R1}) can also be generalized to the case that
the system is not prepared in steady state but an arbitrary
distribution $P({\bf x})$. For $t \geq t'$, we have
\begin{eqnarray}\label{general}
&&R^{\Omega}_i (t- t')
= \int d {\bf x} \Omega({\bf x}) e^{\hat{L} (t-t')} (- \delta \hat{L}) P({\bf x})  \\
&=& -  \Big [ \langle  \Omega(t) \tilde{F}_k (t') D^{-1}_{ik} (t') \rangle
+ \langle \Omega(t) v_k (t') D^{-1}_{ik} (t') \rangle \Big ] \nonumber
\end{eqnarray}
Choose the observable $\Omega =v_i$ and sum over $i$ from equ.
(\ref{general}), the response function in equal time limit $t=t'$
is:
\begin{eqnarray}\label{ds/st}
&&R^{v_i}_i (t) = \int d {\bf x} v_i ({\bf x}) [- \partial_i P ({\bf x}) ]
%&=& \sum _i \int d {\bf x} j^{SS}_i ({\bf x}) (- \frac{\partial}{\partial x_i}) ln P^{SS}({\bf x}) \nonumber \\
=\int d {\bf x} [ \partial \cdot {\bf j} ({\bf x})]  \ln P({\bf x}) \nonumber \\
&&= \frac{d}{dt} \int d {\bf x} P({\bf x})  \ln P ({\bf x}) = - \dot{S}  \nonumber \\
&&= -  \Big [ \langle  v_i (t) \tilde{F}_k (t) D^{-1}_{ik} (t) \rangle
+ \langle v_i (t) v_k (t) D^{-1}_{ik} (t) \rangle \Big ]
\end{eqnarray}
%while
%\begin{eqnarray}
%R^{v_i}_i (t) &=& \int d {\bf x} v_i ({\bf x}) (- \frac{\partial P ({\bf x})}{\partial x_i})
%%&=& \sum _i \int d {\bf x} j^{SS}_i ({\bf x}) (- \frac{\partial}{\partial x_i}) ln P^{SS}({\bf x}) \nonumber \\
%=\int d {\bf x} [ \partial \cdot {\bf j} ({\bf x})]  ln P({\bf x}) \nonumber \\
%& =& \frac{d}{dt} \int d {\bf x} P({\bf x})  ln P ({\bf x}) = - \frac{d S}{dt}
%%= \int d {\bf x} ({\bf \partial \cdot v}^{SS}) P^{SS}({\bf x}) = \langle {\bf \partial \cdot v}^{SS} \rangle
%\end{eqnarray}
Then, the Gibbs entropy $S= - \int d({\bf x}) P({\bf x}) \ln P ({\bf
x})$ has two parts:
\begin{eqnarray}
\dot{S} &=&   \langle v_i
\partial_i \ln [P({\bf x})] \rangle \nonumber \\
&=& \langle v_i D^{-1}_{ij} v_j \rangle + \langle v_i
D^{-1}_{ij} \tilde{F}_j  \rangle  = e_p -
\dot{S}_m
%&= &- \langle v^{SS}_i D^{-1}_{ij} \tilde{F}_j  \rangle %= -\frac{1}{2} \langle {\bf \nabla \cdot v}^{SS} \rangle
\end{eqnarray}
$e_p \geq 0$ is the average entropy production rate of the system
and $T \dot{S}_m = \langle T\dot{s}_m \rangle$ is the average heat
dissipation in the medium. The rate of heat dissipation in the
medium is $\dot{q} = F_i \dot{x}_j D^{-1}_{ij} = T\dot{s}_m$, where
the exchanged heat $q$ is identified with the increase of entropy
$s_m$ in the medium of temperature $T$ \cite{Seifert1}. $\dot{S}$ links with the gradient of the time
dependent probability distribution: $\nabla P$, which is composed of
two terms. One is from the
 bulk entropy production of the system which links with the flux ${\bf v}$
and the other is from the heat dissipation into the medium (surface)
which links with the driving force ${\bf \tilde{F}}$ \cite{Seifert1}. We see
that the driving force for entropy production is the flux. With the
detailed balance, only the time dependent flux contributes to
entropy production. While without detailed balance, entropy
production has both time dependent and steady state flux
contributions. We would like to separate the contribution of time
dependent and independent entropy production of the system and
relate that to the relaxation of time dependent probability and
steady state flux explicitly. Therefore, if we take the observable
$\Omega =v_i - v_i^{SS}$ and sum over $i$, the response function in
equ. (\ref{general}) with equal time limit $t=t'$ gives:
\begin{eqnarray}
&&\langle v_i \partial_i \ln \Big [P^{SS} ({\bf x}) /P ({\bf x}) \Big ] \rangle = \dot{F}_{free}/T \nonumber \\
&=&\langle v^{SS}_i D^{-1}_{ij} v_j \rangle -\langle v_i D^{-1}_{ij} v_j \rangle = Q_{hk}/T - e_p
\end{eqnarray}
It leads to $T e_p = Q_{hk}  -\dot{F}_{free}$ with free energy $F_{free} = T \langle  \ln \big [ \frac{P ({\bf x})}{P^{SS} ({\bf x})} \big ] \rangle = U - TS$,
the house keeping heat $Q_{hk}=T \langle v^{SS}_i (t) v_j (t) D^{-1}_{ij} (t) \rangle=T \langle v^{SS}_i (t) v^{SS}_j (t) D^{-1}_{ij} (t) \rangle=
T e_p + \dot{F}_{free} = T \dot{S}_m +\dot{U}$ and total energy $U = -T \int d {\bf x}  P({\bf x}) \ln [ P^{SS} ({\bf x})]$, which was given in previous literature \cite{Seifert1,Esposito, Ge}.
The change of the total internal energy is $\dot{U} =T \langle v_i (t) \partial_i \ln [P^{SS} ({\bf x})] \rangle$.
%the internal energy $E({\bf x}) = - T
%ln P^{SS} ({\bf x})$ and total energy $U =  \int d {\bf x} E({\bf
%x}) P({\bf x})$ can be introduced phenomenologically \cite{Seifert1,Esposito, Ge}.
%$\dot{U} = \int d {\bf x} E({\bf x}) \dot{P}({\bf x})$
%Then, using $ln \big [P ({\bf x}) \big ] = ln \big [ \frac{P ({\bf x})}{P^{SS} ({\bf x})} \big ] + ln \big [P^{SS} ({\bf x}) \big ]$
%and equ. (\ref{distri_ndb}): $- \tilde{F}_i ({\bf x})  = v^{SS}_i ({\bf x}) - D_{ij} ({\bf x}) \frac{\partial}{\partial x_j} ln [ P^{SS} ({\bf x}) ]$, equ. (\ref{ds/st}) can be rewritten as:
%In other words, $T e_p =-\dot{F}_{free} + Q_{hk}$.
There are two different origins of the total entropy production
$e_p$. $\dot{F}_{free}$ is from spontaneous non-stationary
relaxation which links with the gradient of relative potential $-
\partial_i \ln \Big [\frac{P ({\bf x})}{P^{SS}
({\bf x})} \Big ] $. $Q_{hk}$ is the driving force necessary to
sustain the non-equilibrium environment, which links with the steady
state flux ${\bf v}^{SS} ({\bf x})$. For the non-equilibrium steady
state, $\dot{F}_{free}=0$. $Q_{hk}$ equals the medium dissipated heat
for maintaining the violation of detailed balance: $Q_{hk} = T
\dot{S}_m  = - T \langle v^{SS}_i D^{-1}_{ij}
\tilde{F}_j \rangle$. For detail balanced cases, $Q_{hk}=0$ and
total entropy production of the system equals the spontaneous
relaxation of free energy $T e_p = - \dot{F}_{free}$. Here we found
that the generalized FDT in the equal time limit $t'=t$ naturally
leads to non-equilibrium thermodynamics with total entropy
production from both non-stationary spontaneous relaxation and
stationary house keeping part. This is our first main result.

In addition, we can relate the non-equilibrium {\it Fokker-Planck
equation} with {\it Abelian Gauge Theory} and internal curved space,
as in {\it Quantum Electrodynamics (QED)}\cite{Peskin}. With the
covariant derivative $\nabla_i = \partial_i - D^{-1}_{ij}
\tilde{F}_j=
\partial_i + A_i$,
%as in {\it Abelian gauge theory},
 {\it Fokker-Planck
equation} can be rewritten as: $D_{ij} ({\bf x}) \nabla_j P ({\bf x}) = j_i ({\bf x})$.
The curvature of internal charge space due to the {\it Abelian gauge field}
$A_i$ is:
\begin{equation}\label{curvature}
R_{ij}=\partial_i A_j -  \partial_j A_i=[\nabla_i, \nabla_j].
\end{equation}
where $[\cdot]$ indicates a commutator of two operators. According to equ.
(\ref{distri_ndb}), for the detailed balance case: ${\bf j}^{SS} = 0$,
$A_i = \partial_i \ln(P^{SS})$ is a pure gradient and the curvature is zero: $R_{ij}=0$
which corresponds to a flat space. While for non-equilibrium
cases, ${\bf A}$ can't be written as a gradient and $R_{ij} \neq
0$ which corresponds to a curved internal space. $R_{ij}$ is gauge invariant tensor:
for a gauge transformation $A_i \rightarrow A_i +
\partial_i \phi$, $R_{ij} \rightarrow R'_{ij} = R_{ij}$.
Furthermore, the probabilistic velocity ${\bf v} ({\bf x},t)$ and
the flux ${\bf j} ({\bf x},t)$ are also related to this
internal curvature as:
\begin{equation}\label{curvature1}
\partial_i (D^{-1}_{jk} v_k) - \partial_j (D^{-1}_{ik} v_k)  = R_{ij}
\end{equation}
or in the case  of constant diffusion coefficient $D_{ij} = D
\delta_{ij}$: $\partial_i v_j - \partial_j v_i  = R_{ij}$.
%\begin{equation}
%\partial_i v_j - \partial_j v_i  = R_{ij} .
%\end{equation}
We noticed that equ. (\ref{curvature1}) is also gauge invariant.
It means if we change $A_i \rightarrow A_i + \partial \phi$,
although $P({\bf x}, t)$, ${\bf v} ({\bf x}, t)$ and
${\bf j} ({\bf x}, t)$ are all changed, equ. (\ref{curvature1}) is
always satisfied with a same curvature $R_{ij}$. Moreover, even
${\bf v} ({\bf x}, t)$ and ${\bf j} ({\bf x}, t)$ depend on the
solution of $P({\bf x}, t)$, they always satisfy equ.
(\ref{curvature1}), either for steady state solutions or time
dependent solutions. So $R_{ij}$ represents a measurement of
internal geometry of the non-equilibrium dynamics. This curvature
of internal space relates to the the heat dissipation in the
medium along closed loop.
 Along any
specific path ${\bf x} (t)$, $T \Delta s_m$ is the heat
dissipation in the medium:
\begin{eqnarray}\label{d_sm}
&&T \Delta s_m ({\bf x}'(t'), {\bf x}(t)) = T\int_{t'}^{t} \dot{s}_m dt   \\
&=& \int_{t'}^{t} D^{-1}_{ij} ({\bf x} (t))  \tilde{F}_j ({\bf x} (t))
\dot{x}_i dt =-\int_{t'}^{t} A_i ({\bf x} (t)) d x_i(t) \nonumber
\end{eqnarray}
Using the {\it Stokes's theorem} and the current definition in
equ. (\ref{current1}), the entropy increase of the medium $\Delta
s_m$ along a close loop ${\it C}$ can be written as:
\begin{eqnarray}\label{sm_R}
&& T \Delta s_m^C = - \oint_C A_i ({\bf x} ) d x_i = - \oint_C D^{-1}_{ij}({\bf x} ) v^{SS}_j ({\bf x} ) d x_i \nonumber \\
&=& -\frac{1}{2} \int_{\Sigma} d \sigma_{ij} R_{ij}
\end{eqnarray}
where $\Sigma$ is the surface of the closed loop ${\it C}$, $d
\sigma_{ij}$ is the an area element on this surface, and $R_{ij}$ is
the curvature due to the gauge field ${\bf A}$. Both the curvature
$R_{ij}$ and the close loop heat dissipation in the medium $T \Delta
s^C_m$ are gauge invariant under gauge transformation $A_i
\rightarrow A_i + \partial_i \phi$. Thus, we related non-equilibrium
dynamics to an internal curved space. The presence of the non-zero
flux destroys the detailed balance, leads to non-zero internal
curvature and a global topological non-trivial phase analogous to
quantum mechanical Berry phase \cite{Wang}. This is our second main
point.

%\begin{equation}\label{distri_db}
%- F_i ({\bf x}) P_{SS} ({\bf x}) + \frac{1}{2} \frac{\partial}{\partial x_j} D_{ij} (\bf x) P_{SS} ({\bf x}) = 0
%\end{equation}
%For general nonequilibrium systems without detail balance,
%\begin{equation}\label{distri_ndb}
%- F_i ({\bf x}) P_{SS} ({\bf x}) + \frac{1}{2} \frac{\partial}{\partial x_j} D_{ij} ({\bf x}) P_{SS} ({\bf x}) = j_i ({\bf x})
%\end{equation}
%or equivalently
%\begin{equation}\label{distri_ndb}
%- \tilde{F}_i ({\bf x}) P_{SS} ({\bf x}) + \frac{1}{2} D_{ij} ({\bf x}) \frac{\partial}{\partial x_j} P_{SS} ({\bf x}) = j^{SS}_i ({\bf x})
%\end{equation}

%In the following, we will see the contribution of this gauge field
%and internal curvature in our generalized FDT.
The medium heat dissipation $\Delta s_m$
in equ. (\ref{d_sm}) plays an important role in the
time irreversibility for non-equilibrium systems  \cite{Morriss, Searles, Seifert1}. We will see it also gives an important contribution
in generalized FDT for non-equilibrium dynamics and such contribution
links with this gauge field and internal curvature .

In the following, we will focus on cases of constant diffusion
coefficients $D_{ij}$ for simplicity. If ${\bf j} = 0$, it is the equilibrium
system with detailed balance, which has time reversal invariant: $\langle \Omega (t) F_j ({\bf x}(t')) \rangle
= \langle F_j ({\bf x}(t)) \Omega (t') \rangle$. Using the {\it
Langevin} equation (\ref{LE}), $\langle F_i ({\bf x}(t)) \Omega (t')
\rangle = \langle [ \dot{x}_i (t)- \xi_i (t)] \Omega (t') \rangle =
\langle \dot{x}_i (t) \Omega (t') \rangle $, since random force will
not correlate with $\Omega$ of previous time ($t > t'$): $\langle
\xi_i (t) \Omega (t') \rangle = 0$. Then, we arrive at:
\begin{eqnarray}\label{FDT}
R^{\Omega}_i (t-t') = - D^{-1}_{ik} \Big [\frac{d}{dt} \langle  x_k(t) \Omega (t') \rangle \Big ]
\end{eqnarray}
In particular, for the operator $\Omega ({\bf x}) = x_j$, we see
\begin{eqnarray}\label{FDT0}
R^{x_j}_i (t-t') = -  D^{-1}_{ik} \Big [\frac{d}{dt} \langle  x_k(t) x_j (t') \rangle \Big ]
\end{eqnarray}
which is the FDT near equilibrium \cite{Deker}.

However, if  the system is in non-equilibrium state, there is no
detailed balance: ${\bf j} \neq 0$.
%In particular, for observable $\Omega ({\bf x}) = F_j ({\bf x})$,
%from equ. (\ref{R1}),we have
%\begin{eqnarray}\label{R2}
%R^{F_j}_i (t-t') = -   \Big [ \langle  F_j (t) F_k (t') \rangle +
%\langle F_j (t)  v^{SS}_k (t') \rangle
%\Big ] D^{-1}_{ik} \nonumber \\
%\end{eqnarray}
%The first term of equ. (\ref{R2}) represents the contribution by
%correlations of drift forces themselves analogous to the equilibrium
%system. The second term is the contribution of the flux only for
%non-equilibrium dynamics.
%Furthermore, if $D_{ij} = D \delta_{ij}$, equ. (\ref{R2}) provides a suitable way to define a constant temperature for nonequilibrium
%systems as:
%\begin{eqnarray}\label{T}
%T \equiv D = -  \frac{\langle  F^0_i (t) F_j^0 (t') \rangle + \langle F^0_i (t)  \frac{j^0_j}{P^0_{SS}} (t') \rangle}{R_{F^0_i, j} (t-t') }
%\nonumber \\
%\end{eqnarray}
We are often more interested in the direct observable $x_i$ and a
FDT as the form of equilibrium case as in equ. (\ref{FDT0}), in
which we can split out the correlation $\langle  x_k(t) x_i (t')
\rangle$. Without detailed balance, the system is time irreversible:
$\langle \Omega (t) F_j ({\bf x}(t')) \rangle \neq \langle F_j ({\bf
x}(t)) \Omega (t') \rangle$. According to the {\it Fluctuation
theorem} \cite{Morriss, Searles, Seifert1}, we have
\begin{eqnarray}
\ln \frac{P^{SS} ({\bf x'}) \tilde{P}({\bf x}, t|{\bf x'}, t')}{P^{SS} ({\bf x}) \tilde{P}({\bf x'}, t|{\bf x}, t')} = \Delta s_m +
\ln \frac{P^{SS} ({\bf x'})}{P^{SS} ({\bf x})}
\end{eqnarray}
with $\tilde{P}({\bf x}, t|{\bf x'}, t')$ ($\tilde{P}({\bf x'}, t|{\bf x}, t')$) the probabilities of a forward (backward) path.
We define
$\langle \Omega (t) F_i ({\bf x}(t')) \rangle - \langle F_i ({\bf x}(t)) \Omega (t') \rangle
= \int d {\bf x} d {\bf x'} \Omega ({\bf x}) F_i ({\bf x'}) A({\bf x},{\bf x'}, t-t')$
with
\begin{eqnarray}\label{A}
&& A({\bf x},{\bf x'}, t-t')  \\
&=&  P^{SS} ({\bf x'}) P({\bf x}, t|{\bf x'}, t') - P^{SS} ({\bf x}) P({\bf x'}, t|{\bf x}, t') \nonumber \\
&=& P^{SS} ({\bf x'}) \int D[{\bf x}]  \tilde{P} ({\bf x}, t|{\bf x'}, t') \Big(1 -  \frac{P^{SS} ({\bf x})}{P^{SS} ({\bf x'})} e^{-\Delta s_m}\Big ) \nonumber
\end{eqnarray}
$D[{\bf x}]$ is the path integral from ${\bf
x}'(t')$ to ${\bf x} (t)$. Then, we get
\begin{eqnarray}\label{FDT1}
R^{\Omega}_i (t-t') &=& -  D^{-1}_{ik} \Big [\frac{d}{dt} \langle  x_k(t) \Omega (t') \rangle \Big ]  \nonumber \\
&& - D^{-1}_{ik}  \int d {\bf x} d {\bf x'} \Omega ({\bf x}) F_k ({\bf x'}) A({\bf x},{\bf x'}, t-t')  \nonumber \\
&& - D^{-1}_{ik} \langle \Omega(t)  v_k^{SS} (t') \rangle
\end{eqnarray}
%Then,
%\begin{eqnarray}
%A({\bf x},{\bf x'}, t-t') &=& P^{SS} ({\bf x}) P({\bf x'}, t|{\bf x}, t') \nonumber \\
%&& \times \Big( \frac{P^{SS} ({\bf x'})}{P^{SS} ({\bf x})} \int D[{\bf x}] (t) e^{\Delta s_m} -1 \Big )
%\end{eqnarray}
For the operator $\Omega ({\bf x}) = x_j$, the response function reads
\begin{eqnarray}\label{FDT2}
R^j_i (t-t') &=& - D^{-1}_{ik} \Big [\frac{d}{dt} \langle  x_k(t) x_j (t') \rangle \Big ] \nonumber \\
&& - D^{-1}_{ik}  \int d {\bf x} d {\bf x'} x_j F_k ({\bf x'}) A({\bf x},{\bf x'}, t-t')  \nonumber \\
&& - D^{-1}_{ik} \langle x_j (t)  v_k^{SS} (t') \rangle
\end{eqnarray}
The first term is similar to the equilibrium case in equ.
(\ref{FDT0}). The last two terms in equ. (\ref{FDT2}) are zero for
detailed balance case. These two terms are related to the internal
curvature due to the gauge field in space, as shown in equ.
(\ref{curvature1}) and (\ref{sm_R}). In equ. (\ref{A}), the factor
$U(x,y) = e^{-\Delta s_m} = e^{\frac{1}{T}  \int_P A_i ({\bf x} )
d x_i }$ is very similar to the {\it Wilson loop} or {\it Wilson
line} in {\it Abelian gauge theory}, with $\int_P$ indicating the
integral for a path from ${\bf x}$ to ${\bf y}$ \cite{Peskin}. It
describes the irreversibility determined by the heat dissipation
in the medium. The function inside the path integral of equ.
(\ref{A}) is $U(x,y) \frac{P^{SS} ({\bf x})}{P^{SS} ({\bf x'})}=
e^{-\Delta q_{hk}/T}$, where $\Delta q_{hk}$ is the housekeeping
heat along a trajectory. It was  proved $\langle e^{-\Delta
q_{hk}/T} \rangle=1$ \cite{Seifert1}. Along a closed loop, $
e^{-\Delta q^C_{hk}/T} = U(x,x)$.
%with $R_{ij}$ the internal curvature defined in equ. (\ref{curvature1}).
Under the gauge transformation, $U(x,y)$ transforms as: $U(x,y)
\rightarrow e^{\phi (x)} U(x,y) e^{-\phi (y)}$. It also satisfies
the differential equation:
\begin{eqnarray}
\dot{x}_i \nabla_i U(x,y) =0
\end{eqnarray}
It means that the gradient of phase factor (Wilson
lines) contribution from the heat dissipation or house
keeping part for non-equilibrium systems is perpendicular to the
dynamics just as the case in the circular motion. The origin of the
non-zero curvature is the non-zero flux which breaks the detailed
balance for non-equilibrium systems.
%A term $U(x,y)$ analogy to the {\it Wilson line} in {\it Abelian gauge theory} describes the in irreversible
%is determined by the heat dissipation in the medium.
This is the third and last main result of the paper.

\acknowledgments{

%\bibliography{DP_PRL.bib}

\begin{thebibliography}{99}

\bibitem{Wang}
J. Wang, L. Xu and E. K. Wang, Proc. Natl. Acad. Sci. {\bf 105},
12271 (2008); J. Wang, K. Zhang, E.K. Wang, J. Chem. Phys. {\bf
133}, 125103(2010).

\bibitem{Kubo}
Kubo, R. 1966. Rep. Prog. Phys. {\bf 29}, 255 (1966).

\bibitem{Deker}
U. Deker and F. Haake, Physical Review A, {\bf 11}, 2043 (1975).

%\bibitem{Kurchan}
%Kurchan, J., Nature. {\bf 433}, 222 (2005).

\bibitem{Cugliandolo}
L. Cugliandolo, J. Kurchan, and L. Peliti., Phys.
Rev. E. {\bf 55}, 3898 (1997).


%\bibitem{McAdams}
%McAdams, H., and A. Arkin, Trends Genet. {\bf 15}, 65 (1999).

%\bibitem{Kaern}
%M. Kaern, W. J. Blake, and J. J. Collins, 2003. Annu. Rev. Biomed. Eng. {\bf 5}, 179 (2003).

%\bibitem{Hasty}
%Hasty, J., D. McMillen, and J. Collins, Nature. {\bf 420} 224 (2002).

%\bibitem{Ptashne}
%M. Ptashne, {\it A Genetic Switch},  2nd ed. (Cell Press and Blackwell Science, Cambridge, MA, 1992).

%\bibitem{Davidson} Davidson, E.H., D.H. Erwin, Science. {\bf 311}, 796 (2006).

%\bibitem{Kholodenko}
%Kholodenko, B. N., Eur. J. Biochem. {\bf 267}, 1583 (2000).

%\bibitem{Sasai}
%M. Sasai and P. G. Wolynes, Proc. Natl. Acad. Sci. USA. {\bf 100}, 2374 (2003);

%\bibitem{Ptashne1}
%M. Ptashne,  {\it Genes and Signals} (Cold Spring Harbor Lab. Press,
%Woodbury, NY, 2002).

%\bibitem{Jacob}
%F. Jacob and J. Monod, J. Mol. Biol. {\bf 3}, 318 (1961).

%\bibitem{Thattai}
%M. Thattai and A. van Oudenaarden, Proc. Natl. Acad. Sci.
%U.S.A. {\bf 98}, 86148619 (2001).

%\bibitem{Stratonovich}
%R. L. Stratonovich, {\it Topics in the Theory of Random Noise} (Gordon and Breach, New York, 1963).

%\bibitem{Gillespie_0}
%T.D. Gillespie, J. Phys. Chem. {\bf 81}, 2340  (1977).

%T.B. Kepler \& T.C. Elston, Biophys. J. {\bf 81}, 3116 (2001).
%Exact stochastic simulation of coupled chemical reactions.
%Kepler TB, Elston TC (2001) Stochasticity in transcriptional regulation: Origins, consequences,
%and mathematical representations. Biophys J 81:3116Ð3136.

%\bibitem{Martin}
%P. Martin, A. J. Hudspeth, and F. J\"{u}licher, Proc. Natl.
%Acad. Sci. U.S.A. {\bf 98}, 14380 (2001).

%\bibitem{Mizuno}
%D. Mizuno, C. Tardin, C. F. Schmidt, and F. C.
%MacKintosh, Science {\bf 315}, 370 (2007).

%\bibitem{Cugliandolo_1}
%L. Cugliandolo, D. Dean, and J. Kurchan, Phys. Rev. Lett.
%{\bf 79}, 2168 (1997).

\bibitem{Hanggi}
P. Hanggi and H. Thomas, Phys. Rep., 88 (1982) 207

\bibitem{Chetrite}
R. Chetrite, G. Falkovich, and K. Gawedzki, J. Stat. Mech.
(2008) P08005; R. Chetrite, Phys. Rev. E {\bf 80}, 051107 (2009).

\bibitem{Hatano}
T. Hatano and S. Sasa, Phys. Rev. Lett. {\bf 86}, 3463 (2001),
T. Harada and S. Sasa, Phys. Rev. Lett. {\bf 95}, 130602 (2005).

\bibitem{Verley}
G. Verley, K. Mallick and D. Lacoste, Europhys. Lett. {\bf 93} 10002 (2011).

\bibitem{Baiesi}
M. Baiesi, C. Maes, and B. Wynants, Phys. Rev. Lett. {\bf 103}, 010602 (2009).

\bibitem{Zannetti}
E. Lippiello, F. Corberi, and M. Zannetti, Phys. Rev. E {\bf 71}, 036104 (2005).

\bibitem{Vulpiani}
U. M. B. Marconi, A. Puglisi, L. Rondoni and A. Vulpiani, Phys. Rep., {\bf 461}, 111 (2008).

\bibitem{Okabe}
Y. Okabe, Y. Yagi, and M. Sasai, J. Chem. Phys. {\bf 127}, 105107 (2007).

\bibitem{Lu}
T. Lu, J. Hasty, and P. G. Wolynes, Biophys. J. {\bf 91}, 84 (2006).

\bibitem{Gawedzki}
J. R. Gomez-Solano, A. Petrosyan, S. Ciliberto, R. Chetrite, and K. Gawedzki, Phys. Rev. Lett., {\bf 103}, 040601 (2009).

\bibitem{Bechinger}
V. Blickle, T. Speck, L. Helden, U. Seifert, and C. Bechinger, Phys. Rev. Lett., {\bf 96}, 070603 (2006).

\bibitem{Maes}
J. R. Gomez-Solano, A. Petrosyan, S. Ciliberto and C. Maes, J. Stat. Mech. {\bf 2011}, P01008 (2011).

\bibitem{Seifert}
U. Seifert and T. Speck, Europhys. Lett. {\bf 89}, 10007 (2010).

\bibitem{Prost}
J. Prost, J.-F. Joanny, and J. M. R. Parrondo, Phys. Rev. Lett. {\bf 103}, 090601 (2009).


\bibitem{Peskin}
M. E. Peskin and D. V. Schroeder, {\it An Introduction to Quantum Field Theory}, (Addison-Wesley Publishing Company, 1995).

\bibitem{Ge}
G. Hao and H. Qian, Phys. Rev. E. {\bf 81}, 051133 (2010).


\bibitem{Seifert1}
U. Seifert, Phys. Rev. Lett. {\bf 95}, 040602 (2005);
T. Speck and U. Seifert, J. Phys. A: Math. Gen. {\bf 38}, L581 (2005).


\bibitem{Esposito}
M. Esposito and C. Broeck, Phys. Rev. E. {\bf 82}, 011143 (2010).

\bibitem{Morriss}
D. J. Evans, E.G.D. Cohen and G.P. Morriss, Phys. Rev. Lett. {\bf 71}, 2401-2404 (1993).

\bibitem{Searles}
D. J. Evans and D. J. Searles, Phys. Rev. E {\bf 50}, 1645¨C1648 (1994).

\bibitem{Evans}
G.M. Wang, E.M. Sevick, E. Mittag, D.J. Searles and D. J. Evans, Phys. Rev. Lett. {\bf 89}, 050601 (2002).

\bibitem{Andrieux}
D. Andrieux and P. Gaspard, J. Stat. Mech. P02006, (2007).


%\bibitem{Crooks}
%G. E. Crooks, Phys. Rev. E {\bf 60}, 2721 (1999).

\bibitem{Szabo}
G. Hummer and A. Szabo, Proc. Natl. Acad. Sci. {\bf 98}, 3658 (2001).

\bibitem{Jarzynski}
C. Jarzynski Phys. Rev. Lett. {\bf 78}, 2690 (1997).

%\bibitem{Gardiner}
%C.W. Gardiner, {\it Handbook of Stochastic Methods for Physics, Chemistry and the Nature Sciences},  3rd ed. (Springer, NY, 2004).


%\bibitem{Rosenfeld}
%N. Rosenfeld, J. W. Young, U. Alon, P. S. Swain, and M. B. Elowitz,
%Science {\bf 307}, 1962 (2005).

%\bibitem{Sigal}
%A. Sigal, R. Milo, A. Cohen, N. Geva-Zatorsky, Y. Klein, Y. Lirion, N.
%Rosenfeld, T. Danon, N. Perzov, and U. Alon, Nature (London) {\bf 444}, 643
%(2006).

%\bibitem{Gillespie}
%T. D. Gillespie, J. Phys. Phys. {\bf 81}, 2340 (1977).

%\bibitem{McKane}
%A.J. McKane, J.D. Nagy, T.J. Newman, \& M.O. Stefanini, J. Stat. Phys. {\bf 128} 165 (2007);
%D. Lepzelter, H. Feng, \& J. Wang, Chem. Phys. Lett. {\bf 490} 216 (2010).

%
%\bibitem{Hong}
%K.H. Kim, H. Qian and H.M. Sauro, arXiv:0805.4455 (2008).

%\bibitem{Alon}
%U. Alon, {\it An Introduction to Systems Biology: Design Principles of Biological Circuits},  1st ed. (CRC Press, Taylor \& Francis Group, London, 2006).

%
%\bibitem{Walczak}
%A.M. Walczak, J. Onuchic, \& P.G. Wolynes, Proc. Natl. Acad. Sci. USA. {\bf 102}, 18926 (2005).

\end{thebibliography}

\end{document}